\documentclass[a4paper,11pt]{article}
\pdfoutput=1 

\usepackage{jinstpub} 

\title{Using Open Source EDA Tools in ASICs for HEP: A Mixed Comparison}

\author[a]{Édney M. V. Freitas,}
\author[a]{Nicolas Guimarães,}
\author[a]{Rafael Maria,}
\author[a, 1]{Felipe Costa,\note{Corresponding author.}}
\author[a]{Guilherme Milani,}
\author[a]{Bruno Sanches}
\author[a]{and Wilhelmus Van Noije}

\affiliation[a]{Escola Politécnica da Universidade de São Paulo,\\Avenida Prof. Luciano Gualberto, 158, trav. 3, 05508-010 São Paulo, Brazil}

\emailAdd{felipe.w.costa@usp.br}

\abstract{This work compares open-source electronic design automation tools with a commercial environment using three representative integrated circuit blocks in the IHP 130 nm open PDK: a common-mode noise filter, a finite-state machine, and a voltage-controlled oscillator. The study reports design effort and quality of results for digital logic, including area, power, and timing closure, and examines analog layout feasibility. For the finite-state machine at 50 MHz, the open-source flow reached 0.029 mm$^2$ (post-layout) and 4.37 mW (estimated) with 828 standard cells, whereas the commercial flow achieved 0.019 mm$^2$ and 2.00 mW with 497 cells, corresponding to increases of 53\% in area and 118\% in power. The common-mode noise filter totals 1.879 mm$^2$ with 1703 flip-flops at 50 MHz. The voltage-controlled oscillator occupies 0.0025 mm$^2$ and achieves a simulated maximum oscillation frequency of 2.65 GHz. The contribution is a side-by-side quantification of quality of results across digital and analog blocks in the IHP open PDK. The results indicate that open-source tools are viable for early prototyping, training, and collaboration, while commercial flows retain advantages in automation and quality of results when strict targets on power and area or precision analog layout are required.}


\notoc

\proceeding{TWEPP 2025 Topical Workshop on Electronics for Particle Physics\\
  October 6–10, 2025\\
  Rethymno, Crete, Greece}

\begin{document}
\maketitle
\flushbottom

\section{Introduction}
\label{sec:intro}

Application-specific integrated circuits (ASICs) are central to front-end and data-acquisition electronics in high-energy-physics (HEP) experiments, where tight constraints on radiation tolerance, power, area, and latency motivate custom designs~\cite{LHCbUpgrade2024,CMSRun3_2024,HGCAL_Vertical_2024}. At the same time, design costs and license availability can limit iteration speed and collaboration. Open-source electronic design automation (EDA) flows and open process design kits (PDKs) aim to lower these barriers and improve transparency, with manufacturable demonstrations in recent tapeouts~\cite{OpenROAD_DAC19,OpenLane_WOSET20,SkyWaterPDK,IHP_OpenPDK}.

Despite progress in open digital implementation, published side-by-side comparisons with commercial environments in mature technologies relevant to HEP remain limited, especially for quality of results in standard-cell logic and for the feasibility of analog layout. Existing comparative studies focus almost exclusively on digital logic, typically use earlier generations of open-source flows, and do not address analog layout or HEP-specific constraints (e.g. physical-design benchmarks in qflow, RISC-V implementations comparing OpenLane to commercial tools, and FIFO cores evaluated in qflow versus Cadence Encounter)~\cite{Sangani2022,Hesham2021,Acharya2022}. Analog physical design remains challenging; recent academic tools show promise but require careful handling of device matching and symmetry~\cite{MAGICAL_ICCAD19,ALIGN_arXiv20}.

This work compares an open-source flow with a commercial environment in the IHP 130\,nm open PDK across three representative blocks: a common-mode noise filter, a finite-state machine, and a voltage-controlled oscillator. We quantify digital quality of results, tool runtime, and manual iteration effort under a standardized protocol, and present an analog layout case study that relates constraint handling to parasitics and oscillation metrics, with measured and simulated quantities identified explicitly.

The paper is organized as follows. Section~\ref{sec:methods} details the methodology and implementation flows, Section~\ref{sec:blocks} introduces the benchmark blocks, Section~\ref{sec:results} reports quantitative results and design effort, and Section~\ref{sec:conclusions} concludes.

\section{Methods and Flows}
\label{sec:methods}

All designs target the IHP SG13G2 open PDK with consistent process–voltage–temperature (PVT) corners, libraries, and I/O assumptions. Fairness is enforced with identical register-transfer level descriptions or schematics and a shared Synopsys Design Constraints (SDC) file. We report standard digital quality of results (post-route area, power, timing slack, maximum frequency, cell count, congestion), wall-clock runtime, and the number of manual iterations to timing closure, where a manual iteration is a full place-and-route rerun after edits to constraints, scripts, or floorplan. For the VCO, analog layout effort is given qualitatively in person-hours of constraint-aware placement, routing, and parasitic tuning. Measured data accompany simulations when available. Tool and PDK versions are recorded for reproducibility, following open digital-flow practice~\cite{OpenROAD_DAC19,RePlAce_TCAD2018,TritonRoute_TCAD2021,TritonRouteWXL_TCAD2022}.

\subsection{Digital Flow}
\label{sec:digitalflow}

The open-source pipeline uses Yosys+ABC for logic synthesis and OpenLane/OpenROAD for place-and-route and optimization~\cite{OpenROAD_DAC19,RePlAce_TCAD2018,TritonRoute_TCAD2021,TritonRouteWXL_TCAD2022}. Static timing analysis and power estimation use SAIF or VCD activity from gate-level simulation under the IEEE 1800 waveform standard~\cite{OpenTimer_v2_IEEE_DandT_2021,ieee1800_2023}, and the reported digital power figures rely on this vector-based estimation under identical stimuli in both flows. Physical verification uses the PDK KLayout decks and Netgen for LVS; the commercial baseline mirrors these stages with Genus, Innovus, and Tempus. Both flows share the same 50\,MHz SDC with 0.2 ns of skew and 0.3 ns of transition time, it also includes routing-layer limits, antenna repair, and matched floorplan and utilization. Vector-based power is preferred; any vectorless estimates are explicitly identified. Runs use fixed seeds where available and a uniform runtime script, and timing-accurate SDF simulation was available only in the commercial environment.

\subsection{Analog Flow}
\label{sec:analogflow}

Schematic capture and pre-layout simulation use equivalent environments with identical models, stimuli, and PVT corners (Xschem+Ngspice and Virtuoso+Spectre). The benchmark is a differential current-starved ring-oscillator VCO. Oscillation frequency is taken from steady-state transient segments after startup, $K_{\mathrm{VCO}}$ from the slope of $f_{\mathrm{osc}}(V_{\mathrm{CTRL}})$ near nominal bias, and supply current from the same runs with matched numerical controls. Layout applies the same constraint set in both ecosystems: symmetric and matched devices, common-centroid or interdigitated arrangements where appropriate, and consistent use of guard rings, dummies, shielded differential nets, preferred-direction routing, and local decoupling, reflecting current evidence and challenges in automated analog physical design~\cite{MAGICAL_SiliconProven_2022,ALIGN_DAC2019,ALIGN_ASPDAC2023,ASPDAC2024_PerfDrivenAnalogPD}. Parasitic extraction uses equivalent settings (Magic \texttt{ext}/\texttt{ext2spice} or KLayout-PEX on the open side and vendor LPE on the commercial side) with back-annotated post-layout simulation. To limit bias, we equalize inputs, constraints, PVT corners, floorplan and utilization, routing-layer limits, and power activity across flows, document tool-specific differences, and repeat runs on the same host with fixed seeds when possible, and present measured and simulated data side by side.

\section{Blocks and Implementation}
\label{sec:blocks}

\subsection{Common-Mode Noise (CMN) Filter}
\label{sec:cmn}

\begin{figure}[htbp]
\centering
\begin{minipage}{.32\textwidth}
  \centering
  \includegraphics[width=.9\linewidth]{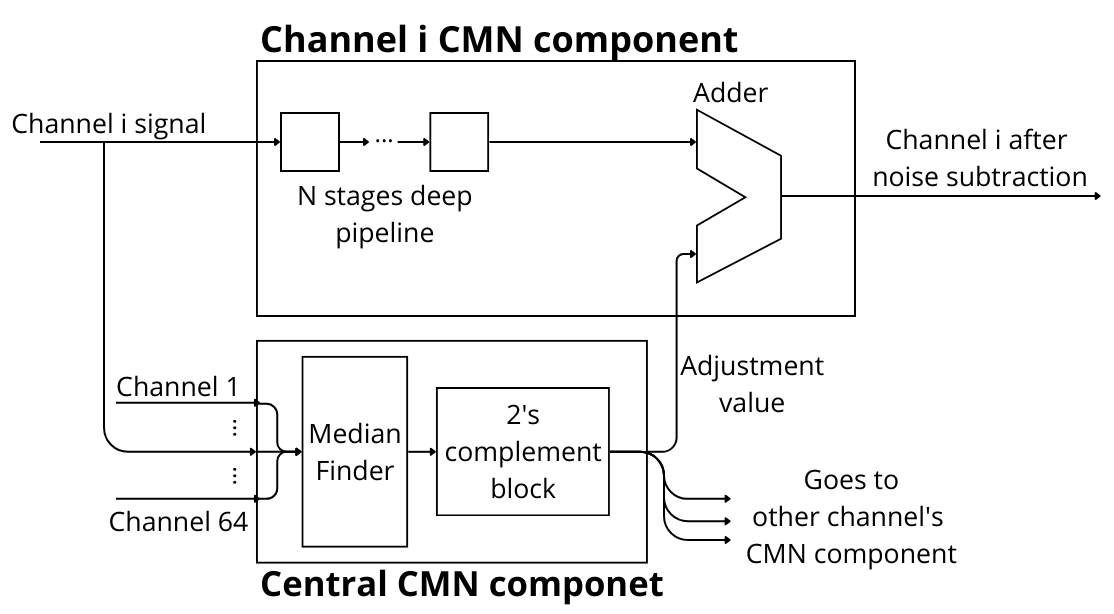}\\[-2pt]
  {\footnotesize (a) CMN filter}
\end{minipage}\hfill
\begin{minipage}{.32\textwidth}
  \centering
  \includegraphics[width=.9\linewidth]{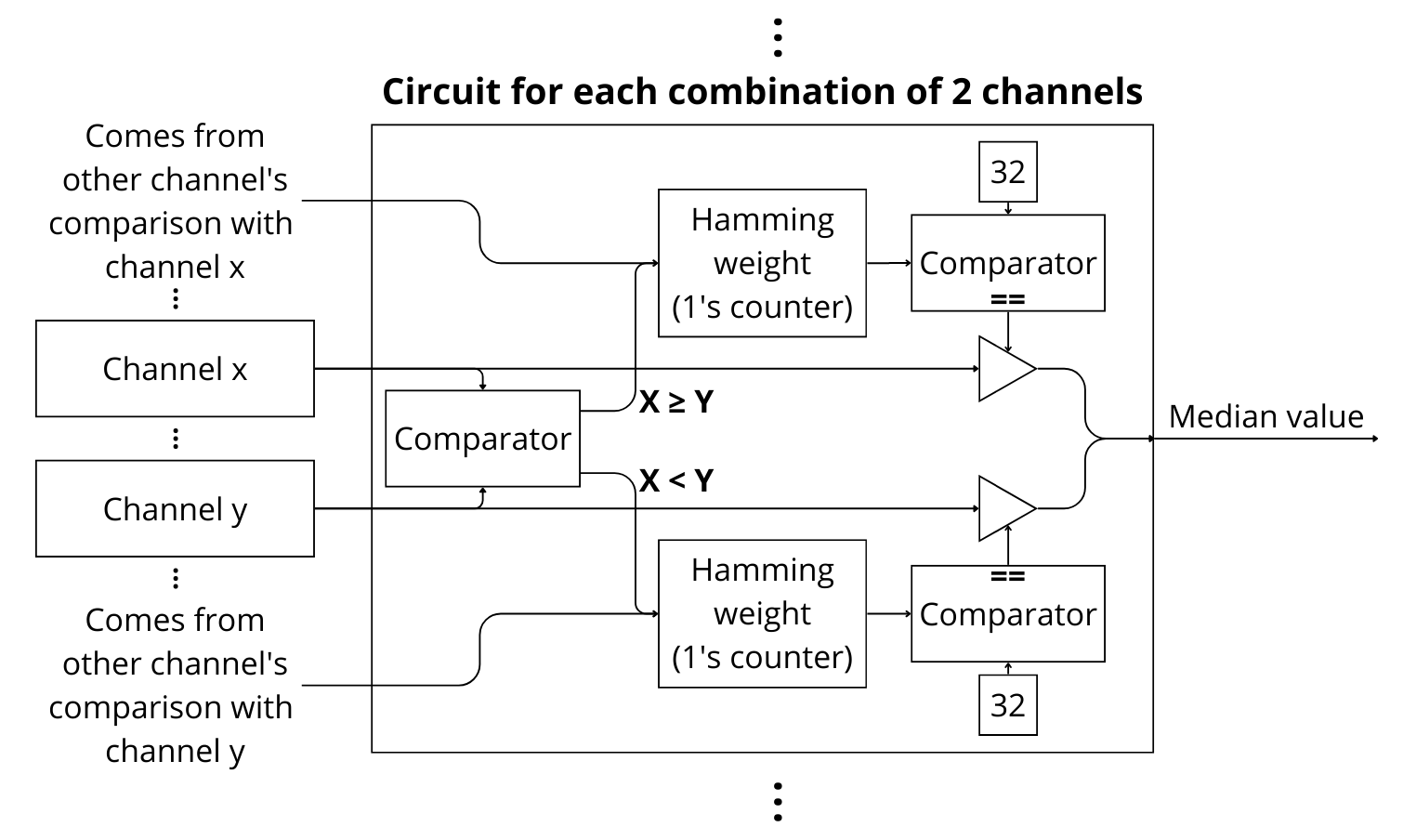}\\[-2pt]
  {\footnotesize (b) CSMF median-finding filter}
\end{minipage}\hfill
\begin{minipage}{.32\textwidth}
  \centering
  \includegraphics[width=.8\linewidth]{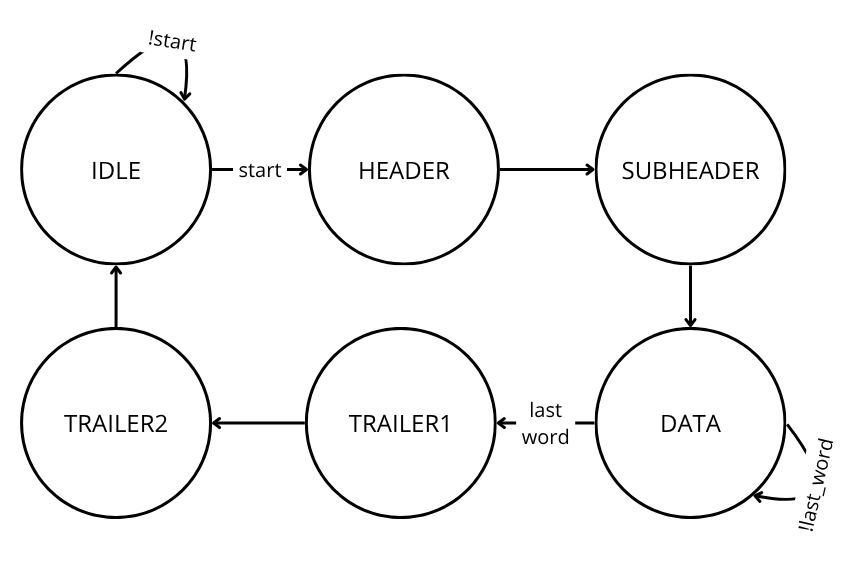}\\[-2pt]
  {\footnotesize (c) Packetizer FSM}
\end{minipage}
\caption{\label{fig:schematics} Schematics of the blocks implemented for comparison.}
\end{figure}

The CMN filter is a digital pre-processor that estimates the common-mode component across $N$ parallel channels and subtracts it from each channel. A hardware median-finding strategy makes the estimate robust to outliers (figure~\ref{fig:schematics}a); a rank-based estimator selects the sample of order $\lceil N/2 \rceil$. The block was developed within the SALSA front-end readout ASIC~\cite{Salsa} under a 50\,MHz pipeline clock and performance, power, and area (PPA) constraints, which motivated a fully combinatorial median solution~\cite{filterscomp}. We use a Combinatorial Sum Median Finder (CSMF) that compares channel pairs in parallel and accumulates per-channel Hamming weights to select the median (figure~\ref{fig:schematics}b), yielding a compact combinatorial benchmark for synthesis and place-and-route.

\subsection{Finite-State Machine (FSM)}
\label{sec:fsm}

The packetizer FSM orchestrates the SALSA output stream~\cite{Salsa}. A six-state controller emits frames with header, subheader, data words, and two trailer words (figure~\ref{fig:schematics}c), while a shift register snapshots internal registers and formats 32-bit words with timing, identification, and integrity fields. Designed under a 50\,MHz clock and PPA constraints, this sequential workload stresses register placement, clock-tree synthesis with timing targets, and hold fixing on short paths, complementing the CMN benchmark.

\subsection{Voltage-Controlled Oscillator (VCO)}
\label{sec:vco}

The analog benchmark is a differential current-starved ring oscillator in the IHP SG13G2 open PDK, with control node $V_{\mathrm{CTRL}}$, differential outputs (\texttt{OUTP}, \texttt{OUTN}), and enable/test pins. Pre-layout simulations share models, stimuli, and PVT corners across environments; $f_{\mathrm{osc}}$ is taken from steady-state transients, $K_{\mathrm{VCO}}$ from the slope of $f_{\mathrm{osc}}(V_{\mathrm{CTRL}})$, and supply current from the same runs. Layout in both ecosystems uses identical constraints: symmetric matched devices in the differential core, common-centroid or interdigitated mirrors and loads, shielded length-matched differential nets, and consistent guard rings, dummies, preferred routing directions, and local decoupling. In the commercial environment, constraint-driven placement and routing enforce these structures, whereas the open-source flow uses manual device grouping, templates, and iterative routing, requiring more tuning cycles to balance parasitics and causing a slightly wider spread in oscillation metrics across PVT runs. Parasitic extraction and post-layout simulation follow the methodology in Section~\ref{sec:analogflow}.

\section{Results}
\label{sec:results}

Regarding the digital flow, the place-and-route tools report power from vector-based gate-level activity, only the core logic site is measured for area, and KLayout scripts are used to count standard cells. Results are shown in Table~\ref{tab:digital_qor}.

Results for the analog flow use the nominal process–voltage–temperature (PVT) corner defined in Section~\ref{sec:methods}. We report median across runs, which are compiled in Table~\ref{tab:vco}.

\begin{table}[htbp]
\centering
\small
\caption{Digital quality of results at 50\,MHz (IHP SG13G2). Power is estimated from gate-level simulation with identical switching activity in both flows.}
\smallskip
\begin{tabular}{|l|l|c|c|c|}
\hline
Block & Flow         & Cell area [mm$^2$] & Power [mW] & Standard cells [count] \\
\hline
CMN (Core)   & Open-source  & 0.343             & 15.8         & 7886                   \\
CMN (Core)   & Commercial   & 0.161             & 4.74         & 5976                   \\
FSM          & Open-source  & 0.029             & 4.37         & 828                    \\
FSM          & Commercial   & 0.019             & 2.00         & 497                    \\
\hline
\end{tabular}
\label{tab:digital_qor}
\end{table}

\begin{table}[htbp]
\centering
\small
\caption{Differential current-starved ring-oscillator VCO (open-source, IHP SG13G2).}
\smallskip
\begin{tabular}{|l|c|}
\hline
Cell area [mm$^2$]                        & 0.0025 \\
Maximum oscillation frequency [GHz]        & 2.65   \\
Lowest reported temperature [$^\circ$C]    & $-269$ \\
\hline
\end{tabular}
\label{tab:vco}
\end{table}

\begin{figure}[htbp]
\centering
\begin{minipage}{.48\textwidth}
  \centering
  \includegraphics[width=.55\linewidth]{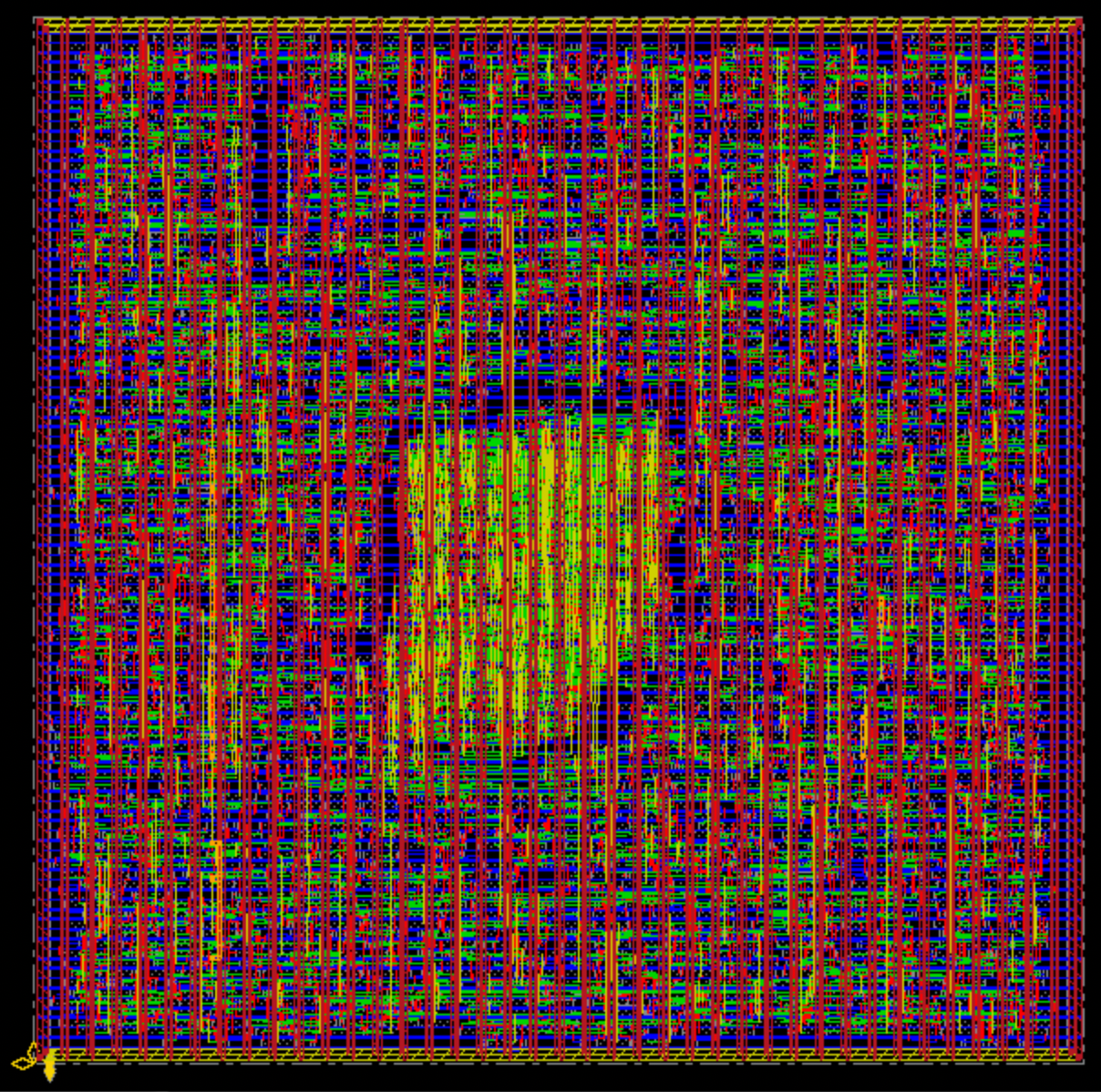}\\[-2pt]
  {\footnotesize (a) CMN core (commercial flow)}\par\smallskip
  \includegraphics[width=.55\linewidth]{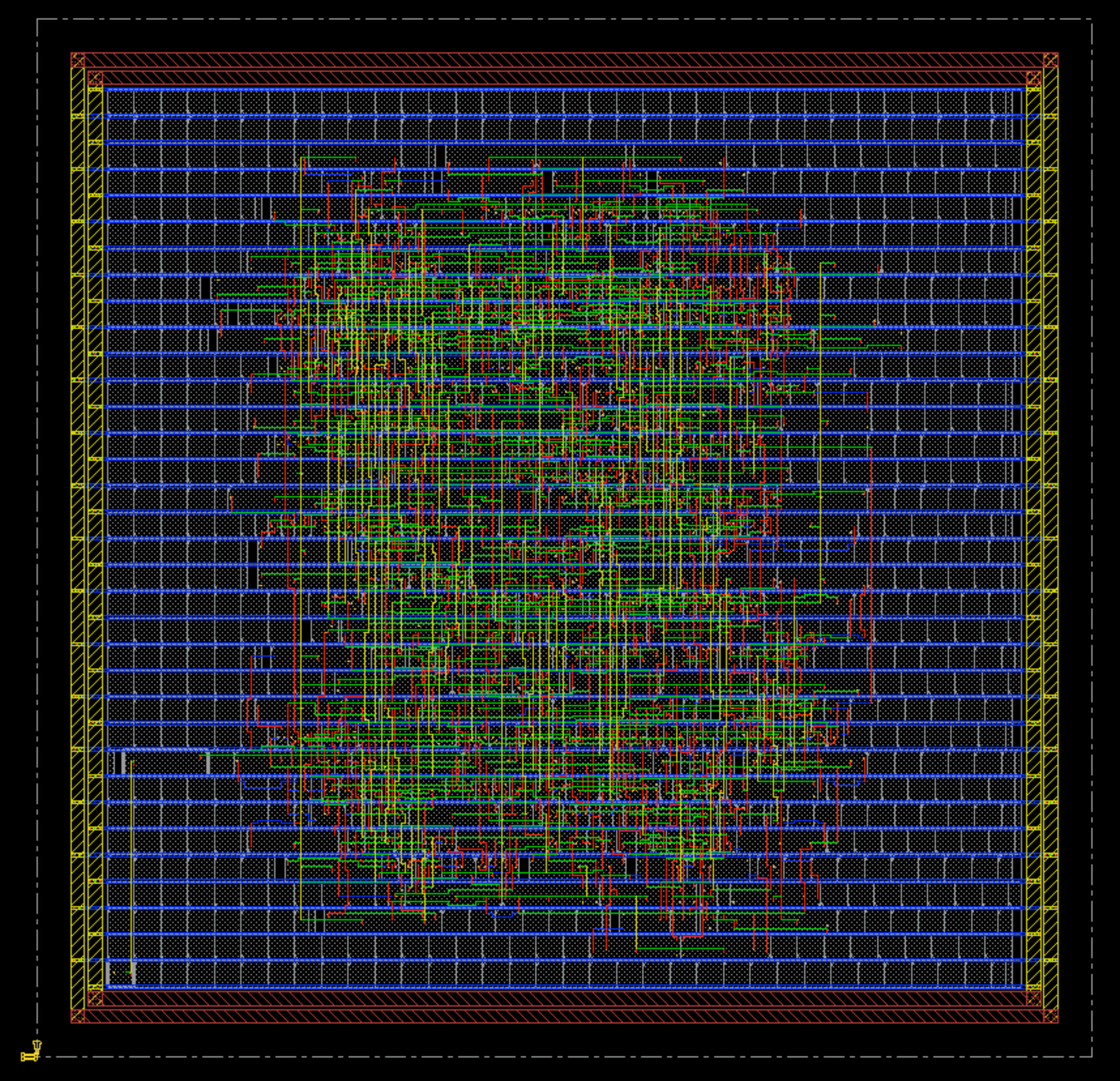}\\[-2pt]
  {\footnotesize (c) FSM (commercial flow)}
\end{minipage}\hfill
\begin{minipage}{.48\textwidth}
  \centering
  \includegraphics[width=.55\linewidth]{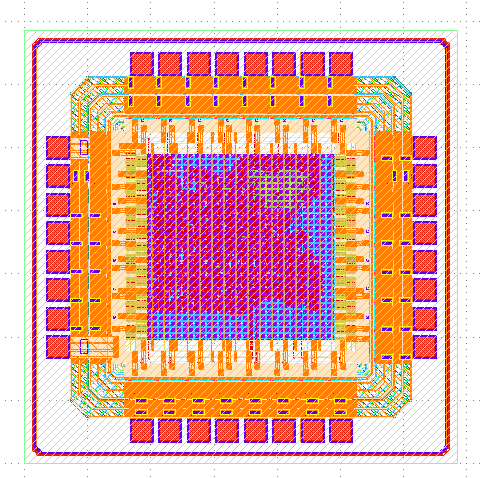}\\[-2pt]
  {\footnotesize (b) CMN padframe view (open-source flow)}\par\smallskip
  \includegraphics[width=.55\linewidth]{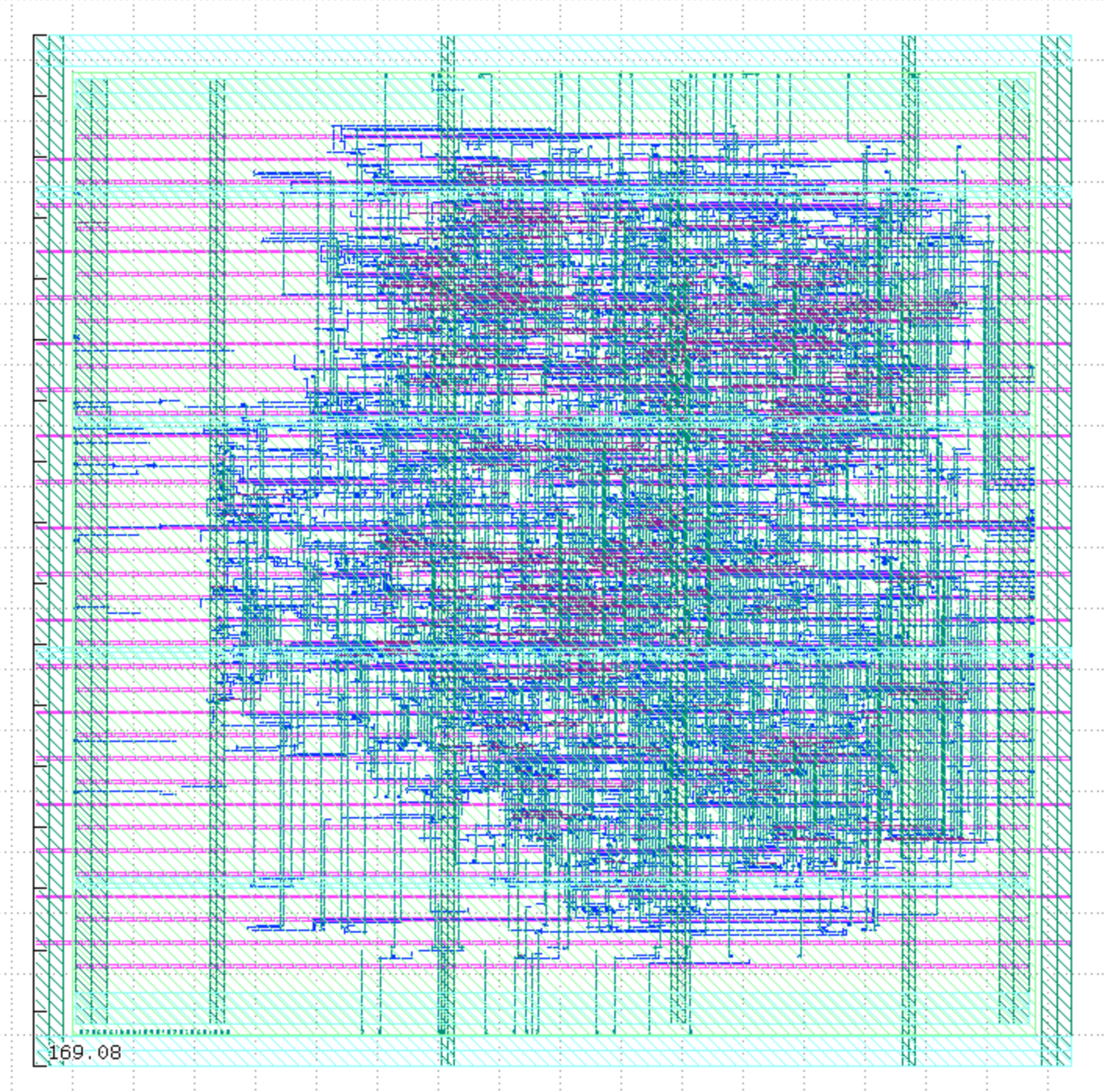}\\[-2pt]
  {\footnotesize (d) FSM (open-source flow)}
\end{minipage}
\caption{\label{fig:dig_layouts}
Digital layouts under identical constraints. Layer colors are tool specific.}
\end{figure}

\begin{figure}[htbp]
\centering
\includegraphics[width=.55\textwidth]{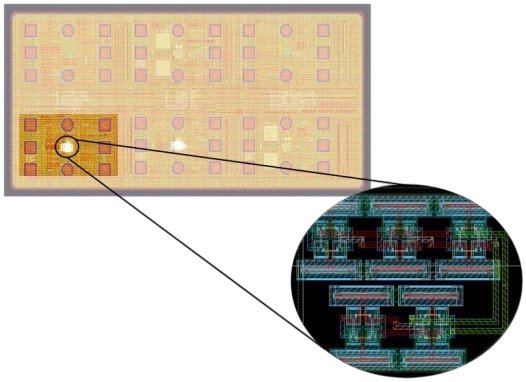}
\vspace{-.5cm}
\caption{\label{fig:vco_layout}
VCO layout with magnified view of the ring core and device tiling. Layer colors are tool specific.}
\end{figure}

\section{Conclusions}
\label{sec:conclusions}

The work compared open-source and commercial flows in the IHP SG13G2 PDK using three representative blocks under identical constraints. The open flow achieved functional digital designs at 50\,MHz and proper analog layouts, while the commercial flow offered smaller area, lower power, and more automated support for constraint-driven layout. The VCO was feasible with open-source tools but required more manual tuning to control parasitics and mismatch. Overall, these free tools are suitable for prototyping, education, and early design exploration, whereas commercial ones remain preferable when strict targets on power, area, or precision analog layout must be met. Future work will expand to more blocks and include silicon measurements.

\acknowledgments
This work was financed, in part, by the São Paulo Research Foundation (FAPESP), Brazil, Grants \#2024/04802-9 and \#2024/06703-8, by CNPQ Grant \#134869/2024-9. This study was financed in part by the Coordenação de Aperfeiçoamento de Pessoal de Nível Superior (CAPES), Brazil – Finance Code 001.

\bibliographystyle{JHEP.bst}
\bibliography{referencesabbrev}
\end{document}